# Absence of transport altermagnetic spin-splitting effect in $RuO_2$

Yu-Chun Wang[1,2], Zhe-Yu Shen[1], Chia-Hsi Lin[1], Wei-Chih Hsu[1], You-Sheng Chen[2], Yi-Ying Chin[3], Akhilesh Kr. Singh[4], Wei-Li Lee[4], Chien-Te Chen[5], Ssu-Yen Huang[1,6&], and Danru Qu[2,6*]

[1] *Department of Physics, National Taiwan University, Taipei 10617, Taiwan*

[2] *Center for Condensed Matter Sciences, National Taiwan University, Taipei 10617, Taiwan*

[3] *Department of Physics, National Chung Cheng University, Chia-Yi 621301, Taiwan*

[4] *Institute of Physics, Academia Sinica, Taipei, 115201, Taiwan*

[5] *National Synchrotron Radiation Research Center, Hsinchu 300092, Taiwan*

[6] *Center of Atomic Initiatives for New Materials, National Taiwan University, Taipei 10617, Taiwan*

[&]syhuang@phys.ntu.edu.tw (S.Y.H.); [*]danru@ntu.edu.tw (D.Q.)

**Abstract**

Altermagnets, which exhibit advantages of both antiferromagnets and ferromagnets, have attracted significant interest recently. Through the transport altermagnetic spin-splitting effect (ASSE), a longitudinal spin polarized or a transverse pure spin current can be generated in the *d*-wave altermagnet upon charge current injection. The ASSE is key experimental features for altermagnet but is often mixed with the spin Hall effect (SHE). In this work, we provide a comprehensive study on the spin-to-charge conversion in epitaxial ruthenium dioxide ($RuO_2$) thin films using the thermal spin current injection from a ferromagnetic insulator yttrium iron garnet (YIG). We conclusively show the absence of ASSE in all our $RuO_2$ films fabricated with three different crystal orientations and deposition methods. We obtain the three independent spin Hall conductivity and spin Hall angle tensor components directly from our experiments. Moreover, we reveal a negative spin Hall angle in $RuO_2$ when it is in proximity to a ferromagnetic insulator YIG but reverses sign when it is next to Py. Our study provides crucial insights into the recent arguments on $RuO_2$, and further advances the understanding of spin-to-charge conversions in emerging materials with low crystal symmetries in general.

## 1. Introduction

Altermagnetism has recently garnered significant attention as a new category of magnetism, alongside ferromagnetism and antiferromagnetism.[1, 2] One prototypical candidate material is ruthenium dioxide ($RuO_2$),[3] which has a rutile crystal structure with space group number 136 ($P4_2$/mnm) and lattice constants $a = b = 4.5$ Å and $c = 3.1$ Å. Early spectroscopic studies, including neutron diffraction[4] and resonant *x*-ray scattering,[5] reveal antiferromagnetic order in $RuO_2$ with Néel vectors aligned along the [001]-direction. Thus, with the $C_2$ magnetic and $C_4$ crystallographic symmetry, $RuO_2$ serves as an ideal metallic *d*-wave altermagnetic candidate.[2] However, recent spectroscopic investigations using muon spin resonance,[6] neutron scattering,[7] and spin- and angle-resolved photoemission spectroscopy[8] have reported the absence of magnetic order in Ru, casting serious doubt on the existence of altermagnetism in this material.

On the other hand, a key advantage of altermagnetic materials is the ability to generate spin current through non-relativistic spin-splitting effects. Accordingly, investigating and understanding the spin-dependent transport in the altermagnetic candidates is crucial for spintronic applications. Pioneer studies in $RuO_2$ have revealed pronounced anisotropic or unconventional spin-charge interconversion.[9-14] Such anisotropic or unconventional spin accumulations are often interpreted as signatures of the altermagnetic spin-splitting effect (ASSE) or its inverse effect (IASSE), which are considered as transport hallmarks of altermagnetism. However, recent spin pumping[15] and terahertz emission[16] experiments suggests the relativistic spin Hall effect (SHE) instead of the ASSE dominates the transport behaviors in $RuO_2$. A more recent spin-splitting torque[17] and spin-splitting magnetoresistance measurements[18-19], however, again consider significant contributions from ASSE-induced spin current. The controversy in the transport study in $RuO_2$ is still not conclusively settled. Since most altermagnets, including $RuO_2$, have inherently low crystal symmetry,[2] a comprehensive understanding of their spin-dependent transport properties is essential to disentangle the origin of the anisotropic and unconventional responses (see Supporting Information S1).

In this work, we provide a comprehensive study on the spin to charge conversion in $RuO_2$. Significantly different from previous reports, we use a ferromagnetic *insulator* yttrium iron garnet (YIG) as capping layer to eliminate charge current complications, and to supply spin current into epitaxially grown $RuO_2$ via the spin Seebeck effect (SSE). To provide a complete comparison, we adopt *three* widely established thin-film deposition techniques: magnetron sputtering, oxide molecular beam epitaxy (oxide MBE), and pulsed laser deposition (PLD) to deposit high-quality epitaxial $RuO_2$ films, on *three* crystal orientations of $TiO_2$ substrates, (100)-, (110)-, and (101)-orientations. Remarkably, we observe robust and anisotropic spin-to-charge conversion in $RuO_2$ regardless of deposition methods and crystal orientations. Through careful analysis, we conclusively show the absence of the transport altermagnetic spin-splitting effects, and a dominant role of anisotropic spin Hall effect in all these $RuO_2$ thin films. We obtain directly from the experiments the three independent spin Hall conductivity and spin Hall angle tensor component for $RuO_2$. Moreover, across the three deposition methods, we consistently observe *negative* spin Hall angles for $RuO_2$ films when it is next to YIG, *opposite* in sign to all the previous reports using Py/ $RuO_2$. Our findings provide critical insight into the recent arguments in $RuO_2$ and offer a framework for studying other altermagnet candidates with low crystalline symmetry.

We begin by illustrating the transverse spin-to-charge conversion in an ideal *d*-wave altermagnet with Néel vectors aligned along the [001]-direction. The *d*-wave spin-splitting bands are depicted in **Figure 1a-d**, where the solid blue and red ellipses represent the energy bands for opposite spin states, indicated by the arrow heads and tails. When a spin current ($J_S$) injects along the [100]-direction, with the spin polarization ($\sigma$) points along the [001]-direction, as illustrated in Figure 1a, the shift of the *d*-wave spin bands (marked by the dotted lines and shaded areas), gives rise to a transverse charge current along the $[0\bar{1}0]$-direction via the IASSE, denoted as $J_{\mathrm{IASSE}}$. In contrast, when $J_{\mathrm{S}}$ is injected along the [110]-direction, or when $\sigma$ is oriented perpendicular to [001], as illustrated in Figure 1b-d, no transverse charge current is generated via IASSE.

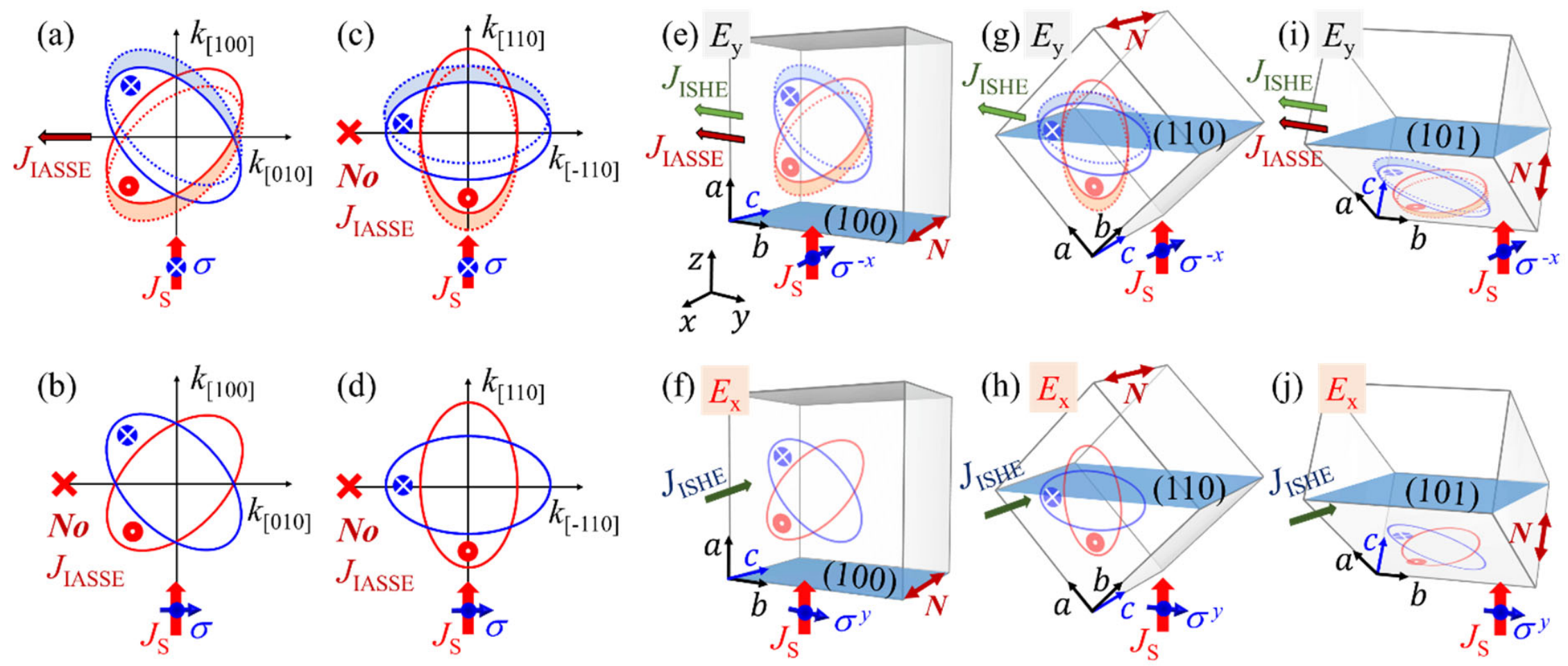

**Figure 1.** A spin current $J_{\mathrm{S}}$ (red arrow) is injected into the *d*-wave spin-splitting bands (red and blue ellipses) along the a) [100]-, b) [100]-, c) [110]-, and d) [110]- directions, with spin polarization $\sigma$ aligned along the a) [001]- , b) [010]-, c) [001]-, and d) $[\bar{1}10]$ -directions, respectively. Arrow tails and heads denote opposite spin states. Solid and dotted ellipses represent the spin-splitting bands before and after spin current injection. The shifted areas of the Fermi surface are shaded in blue and red. Only in a), a transverse charge current is induced by IASSE $J_{\mathrm{IASSE}}$ (dark red arrow). Panels e)-j) illustrate simplified tetragonal unit cells of rutile $RuO_2$, with shaded blue areas representing the (100)-, (110)-, and (101)- crystallographic planes, lying in the *xy*-plane of the measurement coordinate system. The [100]-, [010]-, and [001]-directions are labeled as *a*, *b*, and *c*, respectively. The Néel vector *N* (double arrow) is aligned along the *c*-axis. Spin current is injected along the *z*-axis into the three crystal cuts, with spins aligned along the -*x*- and *y*-axes. A transverse $J_{\mathrm{IASSE}}$ is generated only in the (100)- and (101)-samples in e) and i), but not for the (110)-sample in g), whereas $J_{\mathrm{ISHE}}$ is present in all cases.

As further illustrated in Figure 1e-j, we examine three crystal orientations of $RuO_2$, the (100)-, (110)-, and (101)-planes, which lie in the *xy*-plane of the measurement coordinate system. A spin current $J_{\mathrm{S}}$ is injected vertically into these planes along the *z*-axis. The resulting electromotive force along the *y*-axis is denoted as $E_{\mathrm{y}}$ [Figure 1e, g, i], while that along the *x*- axis is denoted as $E_x$ [Figure 1f, h, j]. Assuming $RuO_2$ is a *d*-wave altermagnet with Néel vector along the *c*-axis, then when $\sigma$ is aligned with the -*x*-axis, as shown in Figure 1e, g, i, only for the (100)- and (101)-, but *not* the (110)-planes, a transverse $J_{IASSE}$ is induced. When $\sigma$ is oriented along the *y*-axis, which is perpendicular to the Néel vectors, as depicted in Figure 1f, h, j, no $E_x$ is generated by IASSE in any of the orientations.

Additionally, an inverse spin Hall effect (ISHE), with an induced charge current $J_{ISHE}$, is expected in all cases shown in Figure 1e-j, due to the sizable spin-orbit coupling in $RuO_2$. Symmetry analysis of the rutile structure with space group No. 136 reveals the coexistence of three independent spin Hall conductivities,[20] which could result in anisotropic spin-to-charge conversion in $RuO_2$. For both the (100)- and (101)- planes, the IASSE and ISHE are mixed and inseparable. But for the (110)-plane, as shown in Figure 1g-h, regardless of the spin orientation, no transverse IASSE is induced. Hence, the (110)-plane plays a crucial role in revealing the anisotropic spin Hall effect in $RuO_2$ and is key to unambiguous distinguishing between the anisotropic ISHE and IASSE experimentally.

## 2. Results and Discussion

The $RuO_2$ layers studied in this work are fabricated on three different orientations, the (100)-, (110)- and (101)-oriented $TiO_2$ substrates, at elevated temperatures of 500 °C using DC sputtering, 350 °C using oxide-MBE, and 650 °C using PLD, and they are denoted as $RuO_2^S$, $RuO_2^{M,}$ and $RuO_2^P$, respectively. The YIG layer is deposited onto the $RuO_2$ layer by radio-frequency (RF) sputtering at room temperature, followed by rapid thermal annealing in an oxygen atmosphere at 800 °C. Magnetization measurements show that after annealing, the YIG layer crystallizes with sizable magnetization [see Supporting Information S2]. For comparison, we also prepare a reference Pt sample, deposited onto the epitaxial YIG film grown on the (111)-oriented gadolinium gallium garnet (GGG) substrate [see Supporting Information S3], as well as a reference permalloy (Py) sample, deposited directly onto the epitaxial $RuO_2$ film at room temperature. All measurements conducted in this work are at room temperature.

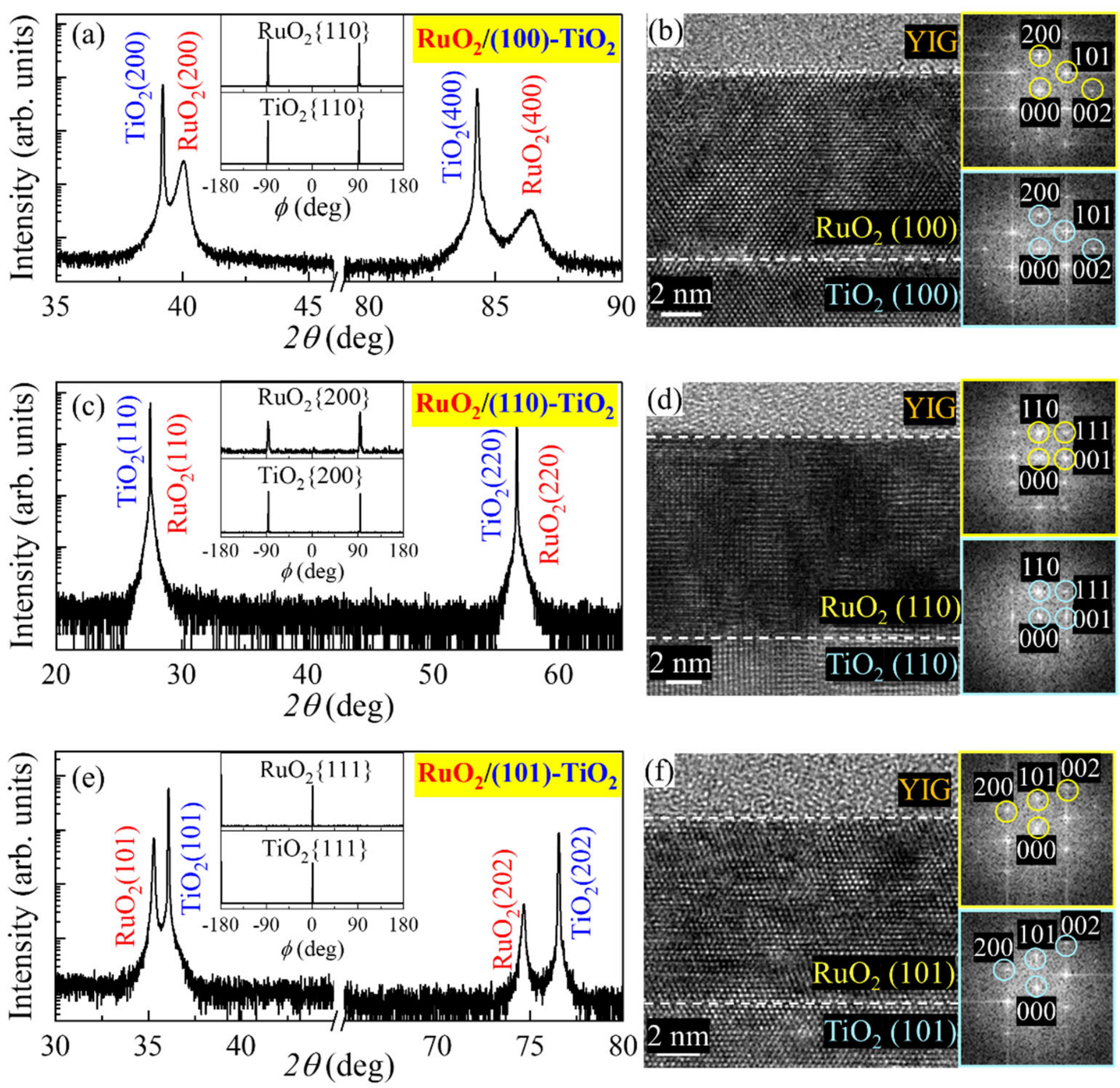

**Figure2.** $2\theta$-scan X-ray diffraction (XRD) of the 100-nm-thick $RuO_2$ grown on the a) (100)-oriented, c) (110)-oriented, and e) (101)-oriented $TiO_2$ substrates, respectively. XRD $\phi$-scan of {110}-planes of (100)-oriented $RuO_2$ and $TiO_2$, {200}-planes of (110)-oriented $RuO_2$ and $TiO_2$, and {111}-planes of (101)-oriented $RuO_2$ and $TiO_2$ are shown in the insets of a), c), and e), respectively. The cross-sectional TEM images of the b) $YIG/RuO_2/TiO_2^{(100)}$, d) $YIG/RuO_2/TiO_2^{(110)}$, and f) $YIG/RuO_2/TiO_2^{(101)}$, exhibiting a structural sharpness across the interfaces. Insets: FFT of the respective $RuO_2$ films (yellow-framed) and $TiO_2$ substrates (blue-framed).

We confirm the epitaxial relationship between the $RuO_2$ layer and the substrate $TiO_2$ using both *X*-ray diffraction spectroscopy (XRD) and transmission electron microscopy (TEM). For the 100-nm thick $RuO_2$ films deposited on the (100)-, (110)-, and (101)- oriented $TiO_2$ substrates, single crystalline peaks are presented for a broad range of the XRD $2\theta$ scan, as shown in **Figure 2a, c, and e**. Further XRD $\phi$-scan in the insets of Figure 2a, c, and e, present two peaks from the $RuO_2$ {110}, {200} and {111} families, which matches nicely with the corresponding peaks for the substrates, revealling the epitaxial relationship. For the thinner 10-nm-thick $RuO_2$ films capped with YIG layers, TEM images disclose sharp interfaces between the YIG and $RuO_2$ layers, and confirms the survival of the epitaxial relationship between $RuO_2$ and $TiO_2$ after the thermal treatment of the YIG layer, as shown in Figure 2b, d, and f. The fast Fourier transform (FFT) in the insets, which show the same patterns for films and substrates further confirms the epitaxy growth of $RuO_2$.

Then we perform the spin Seebeck measurements (SSE) to capture the anisotropic spin-to-charge conversion in $RuO_2$. As shown in **Figure 3a-b**, under a vertical temperature gradient of $\nabla T$ = 13 K $mm^{-1}$, a magnon spin current is driven along the *z*-axis in YIG[21] and injected vertically into the underlying $RuO_2$ layer. The spin current is subsequently converted into a transverse charge current via IASSE or ISHE. The resulting charge accumulations is detected as a voltage ($V$). We estimate the electromotive force ($E$) using $E = V/d$, where $d$ is the distance between the electrodes. The electromotive force obtained along the *x*- and *y*-axes are denoted as $E_x$ and $E_y$, corresponding to those illustrated in Figure 1e-j. For the magnetic field ($H$) angular-dependent measurements, $H$ is rotated within the *xy* plane, with the angle $\phi$ defined relative to the *x*-axis. The $E_y/E_x$ ratio nicely captures the anisotropy of the spin-to-charge conversion in $RuO_2$.

As shown in Figure 3a, for the 15-nm-thick (100)-oriented $RuO_2$ film, a spin current is generated and flows along the *z*-axis, the [100]-direction. With spins aligned parallel (along *x*-axis) and perpendicular (along *y*-axis) to the $[00\bar{1}]$-direction, we observe induced electromotive forces of $E_y$ = - 145 nV $mm^{-1}$ and $E_x$ = - 473 nV $mm^{-1}$, respectively, as shown in Figure 3c. The $E_y/E_x$ ratio is approximately 30%, consistent with that obtained in our previous work.[9]

According to Figure 1e and f, $E_y$ contains contributions from both ISHE and IASSE, whereas $E_x$ arises solely from ISHE. Therefore, if IASSE significantly contributes to the anisotropic spin-to-charge conversion, the voltage ratio for other crystalline orientations, particularly the (110)-oriented $RuO_2$, which has *no* transverse ASSE contribution at all, must be sharply different.

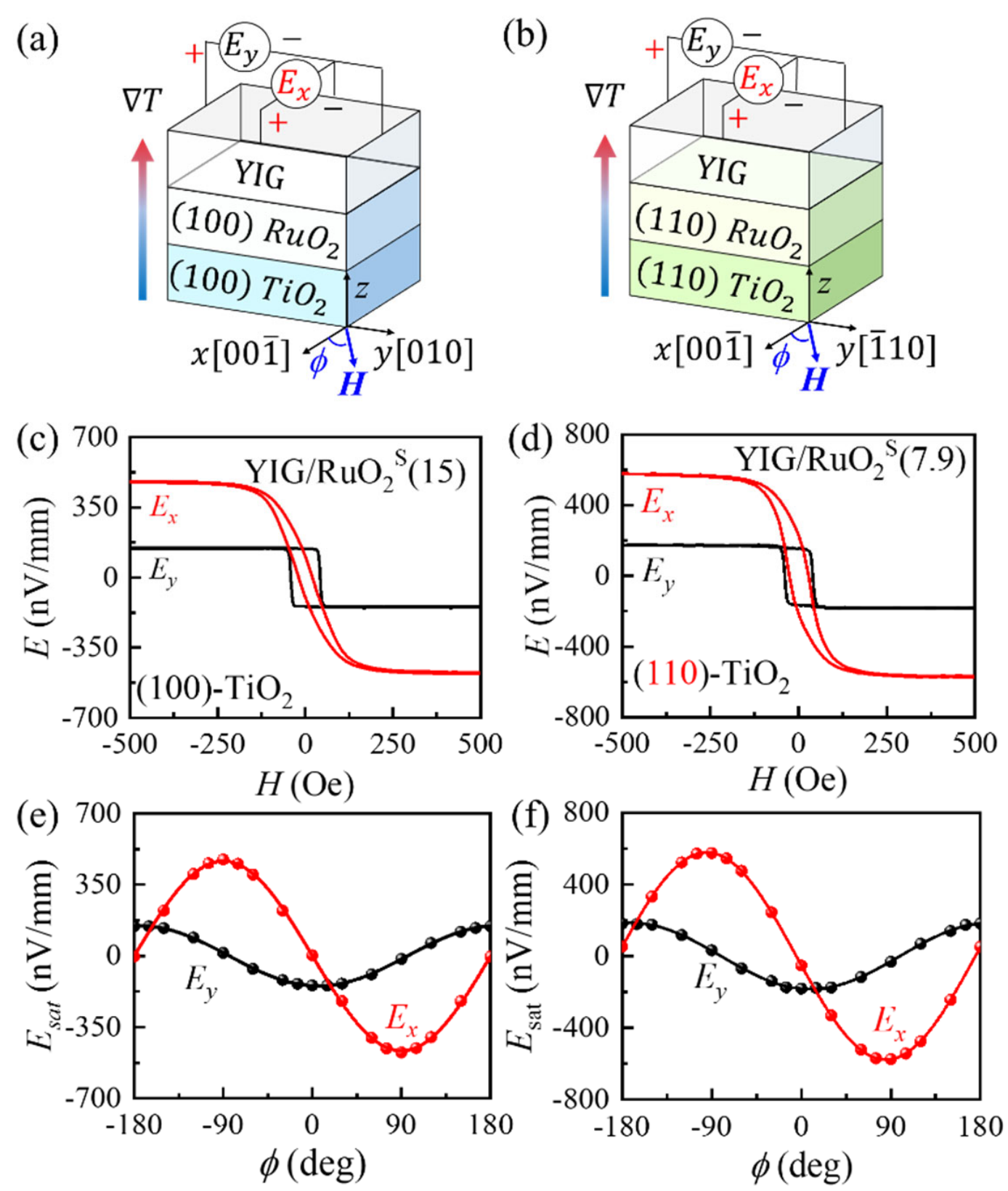

**Figure 3.** Schematic illustrations of the experiment setup for a) (100)- and b) (110)-oriented YIG/$RuO_2$/$TiO_2$ samples. The *x*- and *y*-axes are aligned with the a) [00$\bar{1}$]- and [010]-, b) [00$\bar{1}$]- and [$\bar{1}$10]-crystallographic directions, respectively. The angle $\phi$ denotes the orientation of the external magnetic field *H* relative to the *x*-axis. The spin Seebeck voltage and *H*-angular dependence are measured for the sputtering fabricated c), e) (100)-$RuO_2$ and d), f) (110)-$RuO_2$.

However, as we demonstrated in Figure 3d, for the (110)-oriented $RuO_2$, where the spin current is injected along the [110]-direction and the spin is aligned along the [00$\bar{1}$]- or [$\bar{1}$10]-directions, the voltage signals still exhibit considerable anisotropy with $E_y$ = - 181 nV mm$^{-1}$ and $E_x$= - 574 nV mm$^{-1}$, revealing an $E_y/E_x$ ratio of 30%, nearly *identical* to that observed in the (100)-plane. Without any IASSE contribution, the anisotropic voltage observed in the (110)-$RuO_2$ is *solely* from the ISHE.

From a symmetry point of view,[20] $RuO_2$ with space group No. 136 ($P4_2/mnm$) supports three independent spin Hall conductivity (SHC) tensor components, denoted as $\sigma_{ab}^{c} = -\sigma_{ba}^{c} = A$, $\sigma_{bc}^{a} = -\sigma_{ac}^{b} = B$, and $\sigma_{ca}^{b} = -\sigma_{cb}^{a} = C$, where *a*, *b,* and *c* correspond to the [100] -, [010] -, and [001]- crystal directions, respectively. The altermagnetic spin-splitting conductivity is denoted as $\sigma_{ASSE}$, which accounts for the spin-to-charge conversion via spin-splitting effects. Using a transformation matrix discussed in Supporting Information S4, we derive

$$E_y/E_{x\,(100)} = (A+\sigma_{ASSE})/B \tag{1}$$

for (100)-orientation and,

$$E_y/E_{x\,(110)} = A/B \tag{2}$$

for (110)-orientation.

The nearly identical 30 % ratio for both (100)- and (110)-oriented $RuO_2$ suggests

$$(A+\sigma_{ASSE})/B \approx A/B \approx 30\ \%, \tag{3}$$

which conclusively reveal that $\sigma_{ASSE} \sim 0$. The altermagnetic spin-splitting contribution $\sigma_{ASSE}$ is *absent* in the $RuO_2$ we studied. The observed voltage anisotropy arises *entirely* from the anisotropic spin Hall conductivity, where A/B ≈ 30 %.

As a comparison, for the (101)-oriented film, as illustrated in **Figure 4a**, the spin current flows normal to the film surface, with spins oriented along the [$\bar{1}$01]- and [0$\bar{1}$0]- directions. As shown in Figure 4c, we observe $E_y$ = - 253 nV mm$^{-1}$ and $E_x$ = - 677 nV mm$^{-1}$, yielding a slightly larger $E_y/E_x$ ratio of about 40 %. Using

$$\frac{E_y}{E_{x\,(101)}} = \frac{(A+\sigma_{ASSE})sin^2(\theta_c) + Ccos^2(\theta_c)}{Ccos^2(\theta_c) + Bsin^2(\theta_c)} \tag{4}$$

from Supporting Information S4, where $\theta_c$ = 34.56° denotes the angle between the (001) and (101) plane, A/B ≈ 30 % and $\sigma_{ASSE} \sim 0$, we could extrapolate the relative relationship between the three independent SHC tensors, and obtain C/B ≈ 8 %. These results are summarized in **Table 1**. Notably, the magnetic field ($H$) angular-dependent measurements of $E_y$ and $E_x$, with $H$ rotated in the $xy$ plane, for the (100)-, (110)- and (101)- planes, as illustrated in Figure 3e, f, and Figure 4e, are nicely fit by the cosine and sine functions, with minimum at 0° and 90°, respectively.

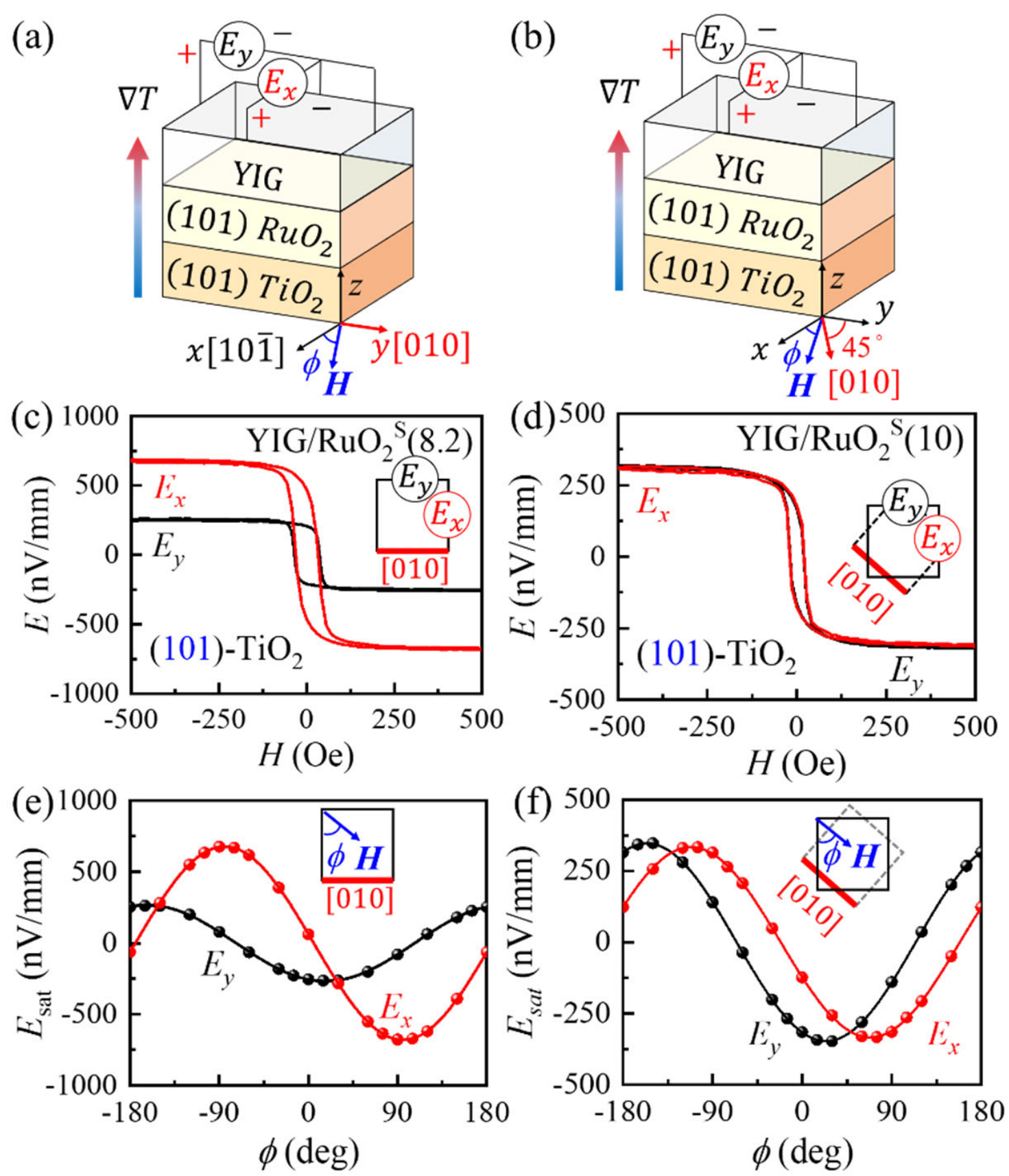

**Figure 4.** Schematic illustrations of the experiment setup for the (101)-oriented YIG/$RuO_2$/$TiO_2$ samples with different in-plane cuts, where $y$-axis is a) aligned with (denoted as regular cut) and b) 45° counter-clockwise-rotated off (denoted as 45°- cut) the [010]- direction, respectively. The angle $\phi$ denotes the orientation of the external magnetic field $H$ relative to the $x$-axis. The spin Seebeck voltage and $H$-angular dependence are measured for the sputtering fabricated (101)-$RuO_2$ films with the c), e) regular cut and d), f) 45°- cut.

To further confirm the absence of ASSE in $RuO_2$, we perform an additional measurement. We employ a different square cut (denote as 45°-cut) of the (101)-oriented sample, as shown in Figure 4b, where the new coordinate $x$ and $y$ axes are rotated 45° in-plane, counter-clockwise off the original coordinates. For the new sample, using A/B = 30 %, C/B = 8 % and $\sigma_{ASSE} = 0$, we expect the voltage ratio $E_y/E_x$ to be 100 %, and voltage minimum locates at 23° and 67°, respectively, for $E_y$ and $E_x$, as discussed in Supporting Information S5. Our experimental results in Figure 4d and 4f show excellent consistency with our prediction. These results further confirm the absence of ASSE contribution.

For the (101)-plane, we also notice that the spin Hall conductivity tensor component $\sigma_{zy}^{z}$ is nonzero (see Supporting Information S5), indicating that a charge current along the $y$-axis generates an unconventional $z$-polarized spins that flow along the $z$-axis, in addition to the conventional $x$-polarized spins. Our calculations yield a $z$-spin to $x$-spin ratio of -0.675, corresponding to an effective spin moment tilted 34° off the $x$-axis towards the -$z$-axis. The $z$-spin can be further utilized to switch a perpendicular magnet.

To verify that the observed anisotropy is independent of fabrication method and extrinsic impurities, we perform the spin Seebeck voltage measurements on $RuO_2$ films grown via PLD and MBE. As shown in **Figure 5a and b**, consistent voltage anisotropy is observed. The $E_y/E_x$ in these $RuO_2$ films remain impressively ~ 30 % for the (100)-plane and ~ 40 % for the (101)-plane. Across a total of 15 samples fabricated by sputtering, PLD and MBE, the averaged $E_y/E_x$ ratios are 31.9 ± 5.6 %, 28.4 ± 4.1 %, and 38.9 ± 5.2 %, for (100)-, (110)-, and (101)-orientations, respectively, as summarized in Figure 5c. Importantly, unlike most studies that report anisotropy using different samples, our work extracts the spin Hall conductivity ratio A/B and C/B within the *same* sample. While the absolute values of A, B, and C may vary by sample, their relative ratios are intrinsic and reproducible, regardless of thickness and fabrication method. The consistency of the $E_y/E_x$ ratios across samples fabricated by different techniques and with varying thicknesses highlight the robustness of the anisotropic spin Hall effect in $RuO_2$, which is governed by its rutile crystal symmetry. The results also consistently suggest the absence of the transport altermagnetic spin-splitting characters in all the $RuO_2$ films we studied.

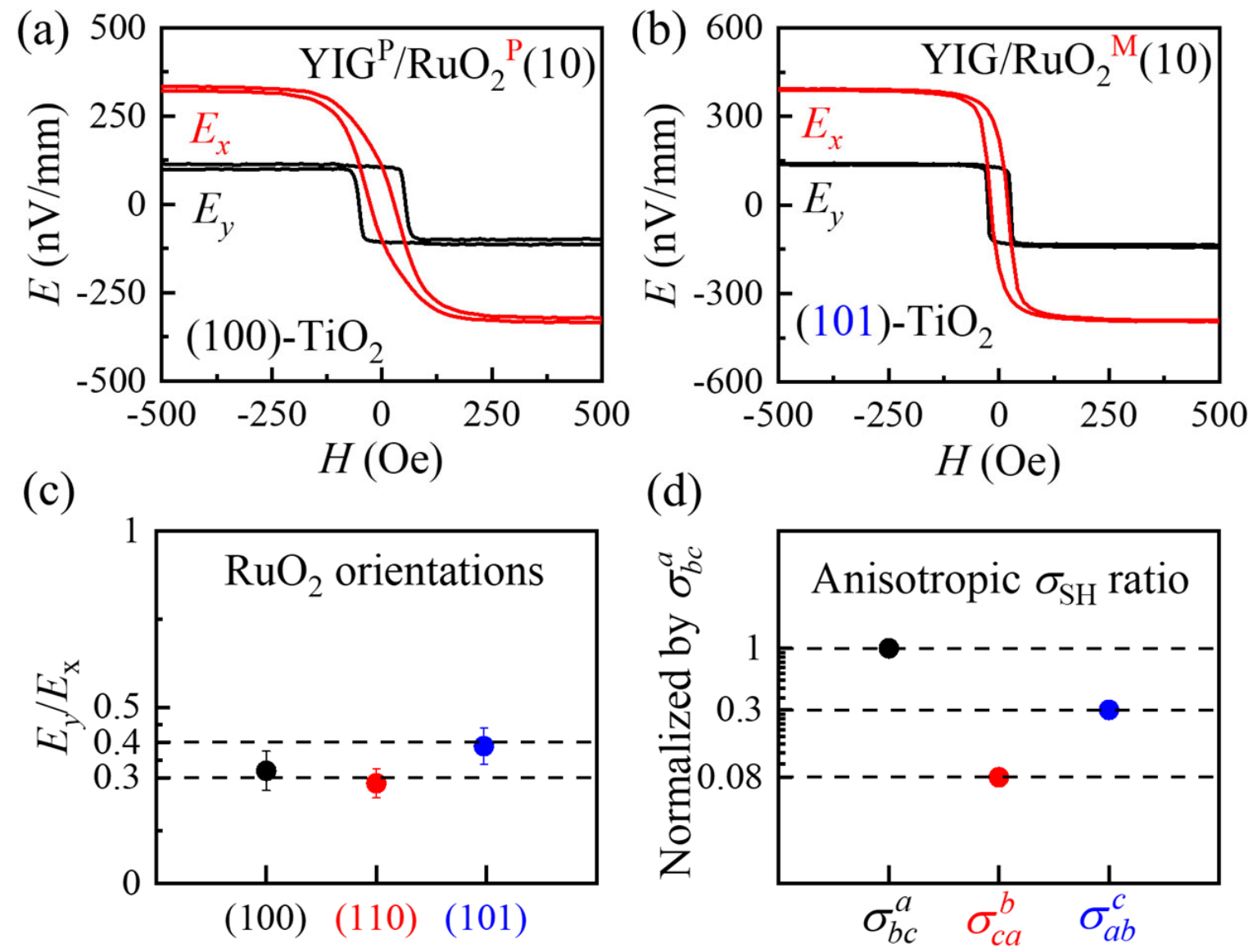

**Figure 5.** Anisotropic voltage signals measured in (a) PLD-fabricated $RuO_2^P$ and (b) MBE-fabricated $RuO_2{}^M$ samples. (c) Summary of orientation-dependent $E_y/E_x$ ratios for all 15 samples examined in this study. (d) Ratios of the anisotropic spin Hall conductivities among the three independent components: $\sigma_{bc}^{a}$, $\sigma_{ca}^{b}$, and $\sigma_{ab}^{c}$.

To understand the magnetic ground state of our $RuO_2$, we perform two independent measurements, including the magnetic field annealing of the YIG/$RuO_2$/$TiO_2$ sample and the X-ray magnetic circular dichroism (XMCD) measurements (see Supporting Information S6-S7). The anisotropic voltage ratio remains unaltered before and after the field annealing process. The XMCD results also reveal undetectable magnetic signals. These results support that the $RuO_2$ in our study is unlikely to be altermagnetic.

Furthermore, to provide a comprehensive study of the spin Hall effect in $RuO_2$, we investigate its spin Hall angle tensor component (${\theta_{\mathrm{SH}}}_{jk}^{i}$), which is often simplified as a single value in other studies. Here, $i$, $j$, and $k$ denotes the direction of spin, spin current and charge current. To identify the sign of $\theta_{\mathrm{SH}}$ for $RuO_2$, we perform a direct comparison with that of Pt.

In Pt, a positive thermal voltage $V$ is observed along the $+y$ direction, when a temperature gradient $\nabla T$ is applied along the $+z$ direction and a magnetic field $H$ is applied along the $-x$ direction, as shown in **Figure 6b**. The sizable electromotive force of $E$ = 1635 nV mm$^{-1}$, as shown in Figure 6d, corresponds to a positive $\theta_{SH} \approx +4$ % for Pt. By contrast, $RuO_2$ under similar experimental conditions [22] (Figure 6a and Supporting Information S8) exhibits a negative $\theta_{SH}$ throughout measurements when YIG is used as a spin current source, as shown in Figure 6c. The negative $\theta_{SH}$ for YIG/$RuO_2$ observed in our study is contrary to prior reports on $RuO_2$ film in proximity to a ferromagnetic metal (FM) layer, such as Py (see Supporting Information S9), but is consistent with the negative $\theta_{SH}$ reported for annealed $RuO_2$ grown on YIG.[23] By further employing the spin pumping measurement, we consistently demonstrate the *opposite* sign in $\theta_{SH}$ for YIG/$RuO_2$ and Py/$RuO_2$ (see Supporting Information S10).

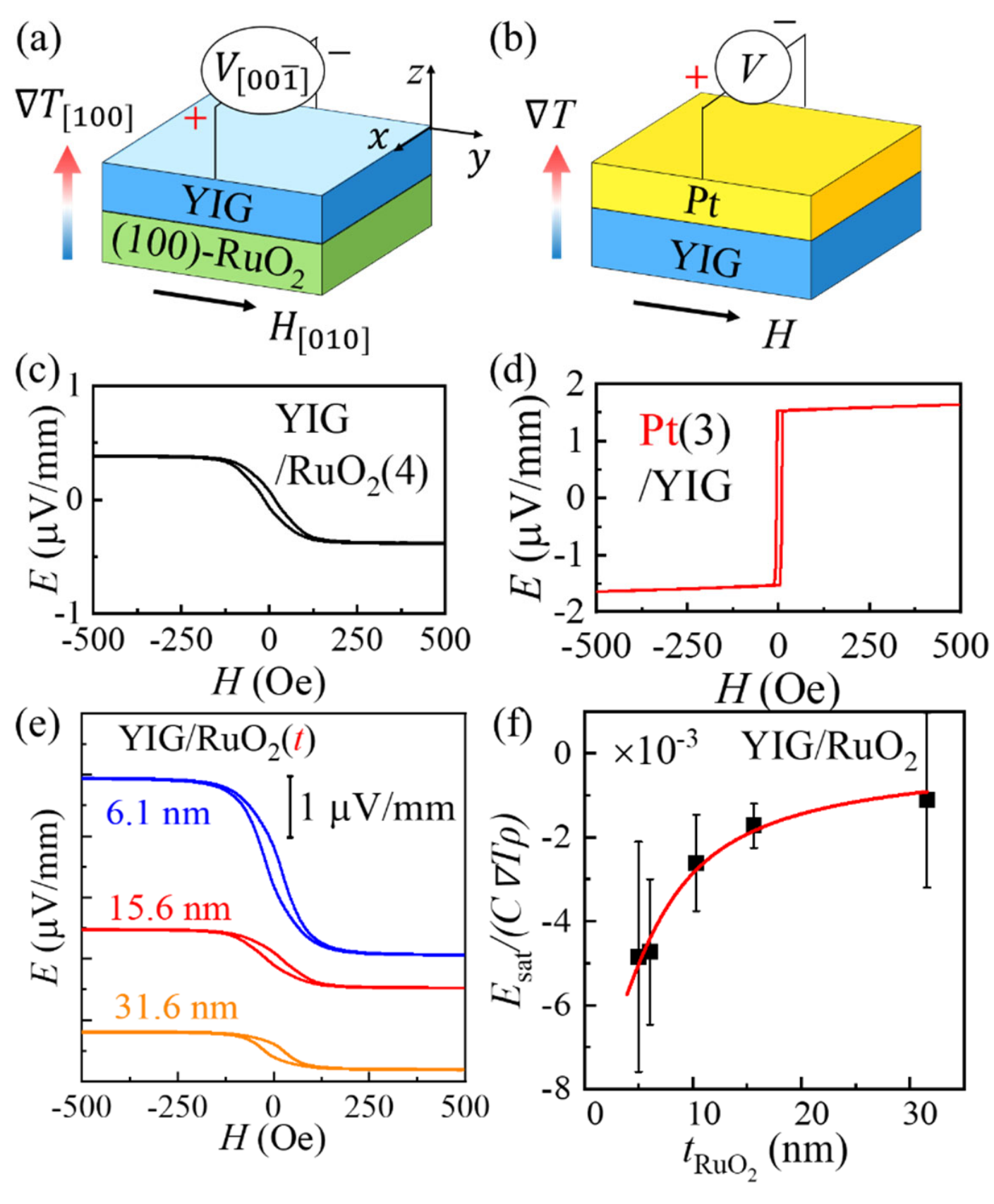

**Figure 6.** Schematic illustrations of the spin Seebeck measurements for (a) YIG/$RuO_2$/$TiO_2$ and (b) Pt/YIG/GGG samples. The spin current injection direction follows the temperature gradient

and thus is the same for (a) and (b), regardless of YIG layer sequence [19]. The spin Seebeck voltages are opposite for the (c) 3.9-nm-thick $RuO_2$ and (d) 3-nm-thick Pt. Thickness dependent (e) spin Seebeck electromotive force and (f) normalized voltage plot for $RuO_2$.

Here we provide a few possibilities that may contribute to the opposite sign for YIG/$RuO_2$ and Py/$RuO_2$. (1) The sizable and positive spin Hall effect in Py.[24-25] This could result in an overall positive $\theta_{\mathrm{SH}}$ for the Py/$RuO_2$ heterostructure. (2) The metallic Ru state at the Py/$RuO_2$ interface, as observed by our HAXPES measurement (see Supporting Information S11), which may significantly modify the spin to charge conversion at the interface. (3) Nontrivial Rashba states at the $RuO_2$ surface,[8] which may be preserved or vanish in proximity to YIG or Py. To conclusively identify the origin of the sign change, a careful and systematic analysis is necessary and awaits further theoretical and spectroscopic insights.

To quantify ${\theta_{\mathrm{SH}}}_{jk}^{i}$, we perform the thickness-dependent ISHE measurement on (100)-oriented $RuO_2$, as shown in Figure 6e. Thicker films show smaller voltages due to spin diffusion. We fit the results in Figure 6f using Equation. S11 in Supporting Information S12, and obtain a ${\theta_{SH}}_{bc}^{a}$ = - (4.0 ± 0.8) % and $\lambda_{\mathrm{sd}}$ = 1.9 ± 0.5 nm. With the estimated resistivity for the bulk $RuO_2$ as 157 μΩcm (see Supporting Information S13), we obtain the anisotropic spin Hall angles and conductivities for $RuO_2$: ${\theta_{SH}}_{bc}^{a} \approx -(4.0 \pm 0.8)\%$, ${\theta_{SH}}_{ca}^{b} \approx -(0.3 \pm 0.06)\%$, ${\theta_{SH}}_{ab}^{c} \approx -(1.2 \pm 0.2)\%$, $\sigma_{bc}^{a} \approx -250 \pm 51\ Scm^{-1}$, $\sigma_{ca}^{b} \approx -19 \pm 3\ Scm^{-1}$, and $\sigma_{ab}^{c} \approx -75 \pm 15\ Scm^{-1}$. These values are summarized in Table 1.

| | | | |
|---|---|---|---|
| Spin Hall Angle | ${\theta_{SH}}_{bc}^{a}$ | ${\theta_{SH}}_{ca}^{b}$ | ${\theta_{SH}}_{ab}^{c}$ |
| (%) | -4.0 ± 0.8 | -0.3 ± 0.06 | -1.2 ± 0.2 |
| Spin Hall Conductivity | $\sigma_{bc}^{a} = -\sigma_{ac}^{b}$ | $\sigma_{ca}^{b} = -\sigma_{cb}^{a}$ | $\sigma_{ab}^{c} = -\sigma_{ba}^{c}$ |
| ($Scm^{-1}$) | -250 ± 51 | -19 ± 3 | -75 ± 15 |
| Anisotropy Ratio | $\sigma_{bc}^{a}/\sigma_{bc}^{a}$ | $\sigma_{ca}^{b}/\sigma_{bc}^{a}$ | $\sigma_{ab}^{c}/\sigma_{bc}^{a}$ |
| (%) | 100 | 8 | 30 |

**Table 1.** Summarization of the anisotropic spin Hall angle, spin Hall conductivity, and anisotropy ratio obtained from the anisotropic spin-to-charge conversion in $RuO_2$.

## 3. Conclusion

In conclusion, we have systematically investigated the anisotropic spin-to-charge conversion in epitaxial $RuO_2$ thin films across different crystal orientations and fabrication methods at room temperature. Most significantly, we show the absence of transport altermagnetic spin-splitting behaviors in all the $RuO_2$ films we studied. Instead, we observe a robust anisotropic spin Hall effect with the spin Hall angle tensor component determined as ${\theta_{SH}}_{bc}^{a} \approx -(4.0 \pm 0.8)\%$, ${\theta_{SH}}_{ca}^{b} \approx -(0.3 \pm 0.06)\%$, ${\theta_{SH}}_{ab}^{c} \approx -(1.2 \pm 0.2)\%$ directly from our experiment. We also show that an unconventional $z$-spin accumulation is expected for the low symmetry (101)-plane which is intrinsic to $RuO_2$ in the absence of magnetic order. Furthermore, we reveal a *negative* spin Hall angle for $RuO_2$ when it is in contact with YIG but changes to positive sign when next to Py. Our study provides critical insights into the recent arguments on $RuO_2$, and advances the understanding of spin-to-charge conversions in emerging altermagnetic materials with low crystal symmetries in general.

Acknowledgement: This work at NTU has been supported by the National Science and Technology Council under Grant No. NSTC 113-2628-M-002-019, NSTC 112-2123-M-002-001, NSTC 113-2112-M-002-039, NSTC 113-2112-M-194-002, and NSTC 113-2124-M-001-011. Center of Atomic Initiative for New Materials (AI-MAT), National Taiwan University, within the Higher Education Sprout Project by the Ministry of Education in Taiwan.